# Monochromatic neutrino beams

J. Bernabeu*, J. Burguet-Castell*, C. Espinoza* & M. Lindroos†

* Universitat de València and IFIC, E-46100 Burjassot, València, Spain
†AB-department, CERN, Geneva, Switzerland

**In the last few years spectacular results have been achieved with the demonstration of non vanishing neutrino masses and flavour mixing[1,2]. The ultimate goal is the understanding of the origin of these properties from new physics. In this road, the last unknown mixing [$U_{e3}$] must be determined. If it is proved to be non-zero, the possibility is open for Charge Conjugation-Parity (CP) violation in the lepton sector. This will require precision experiments with a very intense neutrino source. Here a novel method to create a monochromatic neutrino beam, an old dream for neutrino physics, is proposed based on the recent discovery[3] of nuclei that decay fast through electron capture. Such nuclei will generate a monochromatic directional neutrino beam[4] when decaying at high energy in a storage ring with long straight sections. We also show that the capacity of such a facility to discover new physics is impressive, so that fine tuning of the boosted neutrino energy allows precision measurements of the oscillation parameters even for a [$U_{e3}$] mixing as small as 1°. We can thus open a window to the discovery of CP violation in neutrino oscillations.**

Neutrinos are very elusive particles that are difficult to detect. Even so, physicists have over the last decades successfully studied neutrinos from a wide variety of sources, either natural, such as the sun and cosmic objects, or manmade, such as nuclear power plants or accelerated beams. Spectacular results have been obtained in the last few years for the flavour mixing of neutrinos obtained from atmospheric, solar, reactor and accelerator sources and interpreted in terms of the survival probabilities for the beautiful quantum phenomenon of neutrino oscillations[1,2]. The weak interaction eigenstates $\nu_\alpha$ ($\alpha=e,\mu,\tau$) are written in terms of mass eigenstates $\nu_k$ (k=1,2,3) as $\nu_\alpha = \Sigma_k U_{\alpha k}(\theta_{12}, \theta_{23}, \theta_{13}; \delta) \nu_k$, where $\theta_{ij}$ are the mixing angles among the three neutrino families and $\delta$ is the CP violating phase. Neutrino mass differences and the mixings for the atmospheric $\theta_{23}$ and solar $\theta_{12}$ sectors have thus been determined. The third connecting mixing $|U_{e3}|$ is bounded as $\theta_{13} \leq 10°$ from the CHOOZ reactor experiment[5]. Next experiments able to measure this still undetermined mixing and the CP violating phase $\delta$, responsible for the matter-antimatter asymmetry, need to enter into a high precision era with new machine facilities and very massive detectors. The observation of CP violation needs an experiment in which the emergence of another neutrino flavour is detected rather than the deficiency of the original flavour of the neutrinos. The appearance probability $P(\nu_e \rightarrow \nu_\mu)$ as a function of the distance between source and detector (L) is given by

$$P(\nu_e \rightarrow \nu_\mu) \approx s_{23}^2 \sin^2 2\theta_{13} \sin^2(\frac{\Delta m_{13}^2 L}{4E}) + c_{23}^2 \sin^2 2\theta_{12} \sin^2(\frac{\Delta m_{12}^2 L}{4E}) + \tilde{J} \cos(\delta - \frac{\Delta m_{13}^2 L}{4E}) \frac{\Delta m_{12}^2 L}{4E} \sin(\frac{\Delta m_{13}^2 L}{4E}),$$

(1)

where $\tilde{J} = c_{12} \sin 2\theta_{12} \sin 2\theta_{23} \sin 2\theta_{13}$, $s_{ij}$ and $c_{ij}$ are the corresponding sin and cos functions of $\theta_{ij}$ and $\Delta m_{ij}^2$ the square mass differences. The four measured parameters $(\Delta m_{12}^2, \theta_{12})$ and $(\Delta m_{23}^2, \theta_{23})$ have been fixed throughout this paper to their mean values[6].

The three terms of Eq. (1) correspond, respectively, to contributions from the atmospheric and solar sectors and their interference. As seen, the CP violating contribution has to include all mixings and neutrino mass differences to become observable. Neutrino oscillation phenomena are energy dependent (see Fig. 1) for a fixed distance between source and detector, and the observation of this energy dependence would disentangle the two important parameters: whereas $|U_{e3}|$ gives the strength of the appearance probability, the CP phase acts as a phase-shift in the interference pattern. These properties suggest the consideration of a facility able to study the detailed energy dependence by means of fine tuning of monochromatic neutrino beams. In the CERN Joint Meeting of BENE/ECFA for Future Neutrino Facilities in Europe, the option of a monochromatic neutrino beam from atomic electron capture in $^{150}$Dy was considered and discussed both[4] in its Physics Reach and the machine feasibility. This idea was conceived earlier[7] by the authors and presented together with the beta beam facility. The subsequent analysis showed that this conception could become operational when combined with the recent discovery of nuclei far from the stability line, having super allowed spin-isospin transitions to a giant Gamow-Teller resonance kinematically accessible[3]. Thus the rare-earth nuclei above $^{146}$Gd have a small enough half-life for electron capture processes. In the proposed facility, the neutrino energy is dictated by the chosen boost of the ion source and the neutrino beam luminosity is concentrated at a single known energy.

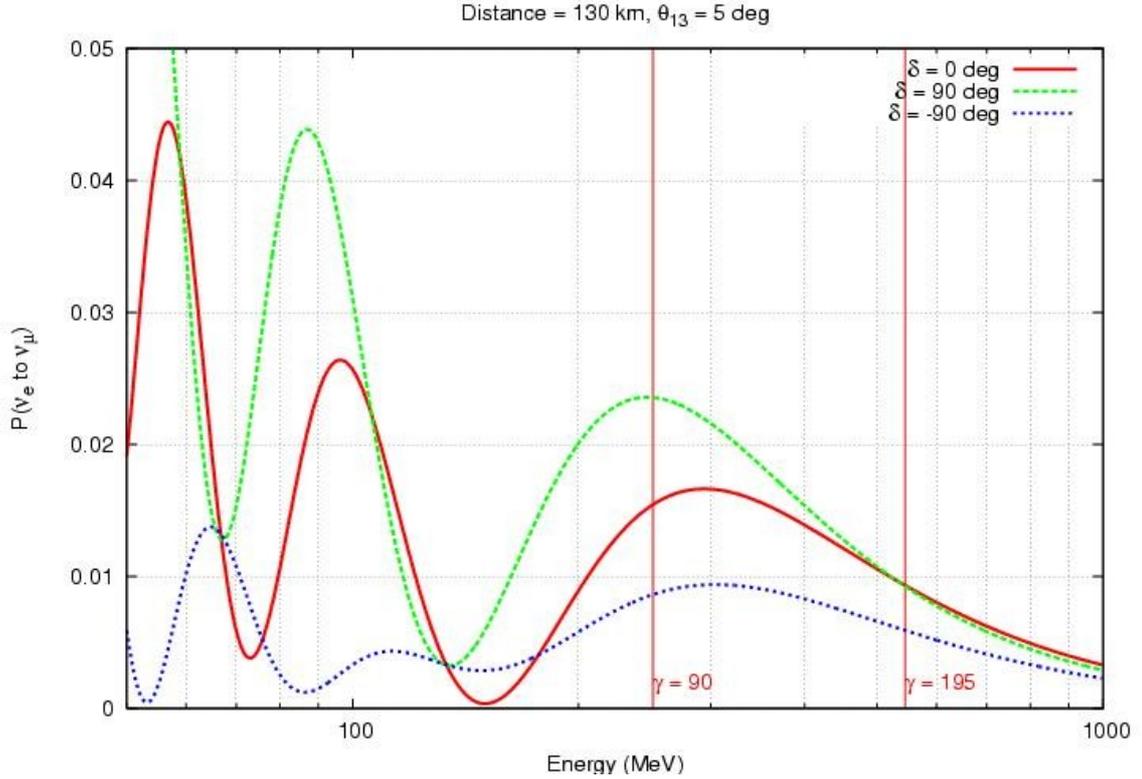

**Figure 1. The appearance probability $P(\nu_e \to \nu_\mu)$ for neutrino oscillations as function of the LAB energy E, with fixed distance between source and detector and connecting mixing. The three curves refer to different values of the CP violating phase δ. The two vertical lines are the energies of our simulation study.**

Electron Capture is the process in which an atomic electron is captured by a proton of the nucleus leading to a nuclear state of the same mass number A, replacing the proton by a neutron, and a neutrino. Its probability amplitude is proportional to the atomic wavefunction at the origin, so that it becomes competitive with the nuclear β+ decay at high Z. Kinematically, it is a two body decay of the atomic ion into a nucleus and the neutrino, so that the neutrino energy is well defined and given by the difference between the initial and final nuclear mass energies ($Q_{EC}$) minus the excitation energy of the final nuclear state. In general, the high proton number Z nuclear beta-plus decay (β+) and electron-capture (EC) transitions are very "forbidden", i.e., disfavoured, because the energetic window open $Q_\beta/Q_{EC}$ does not contain the important Gamow-Teller strength excitation seen in (n,p) reactions. There are a few cases, however, where the Gamow-Teller resonance can be populated (see Fig. 2) having the occasion of a direct study of the "missing" strength. For the rare-earth nuclei above $^{146}$Gd, the filling of the intruder level $h_{11/2}$ for protons opens the possibility of a spin-isospin transition to the allowed level $h_{9/2}$ for neutrons, leading to a fast decay. The properties of a few examples[8] of interest for neutrino beam studies are given in Table 1. A proposal for an accelerator facility with an EC neutrino beam is shown in Fig. 3. It is based on the most attractive features of the beta beam concept[9]: the integration of the CERN accelerator complex and the synergy between particle physics and nuclear physics communities.

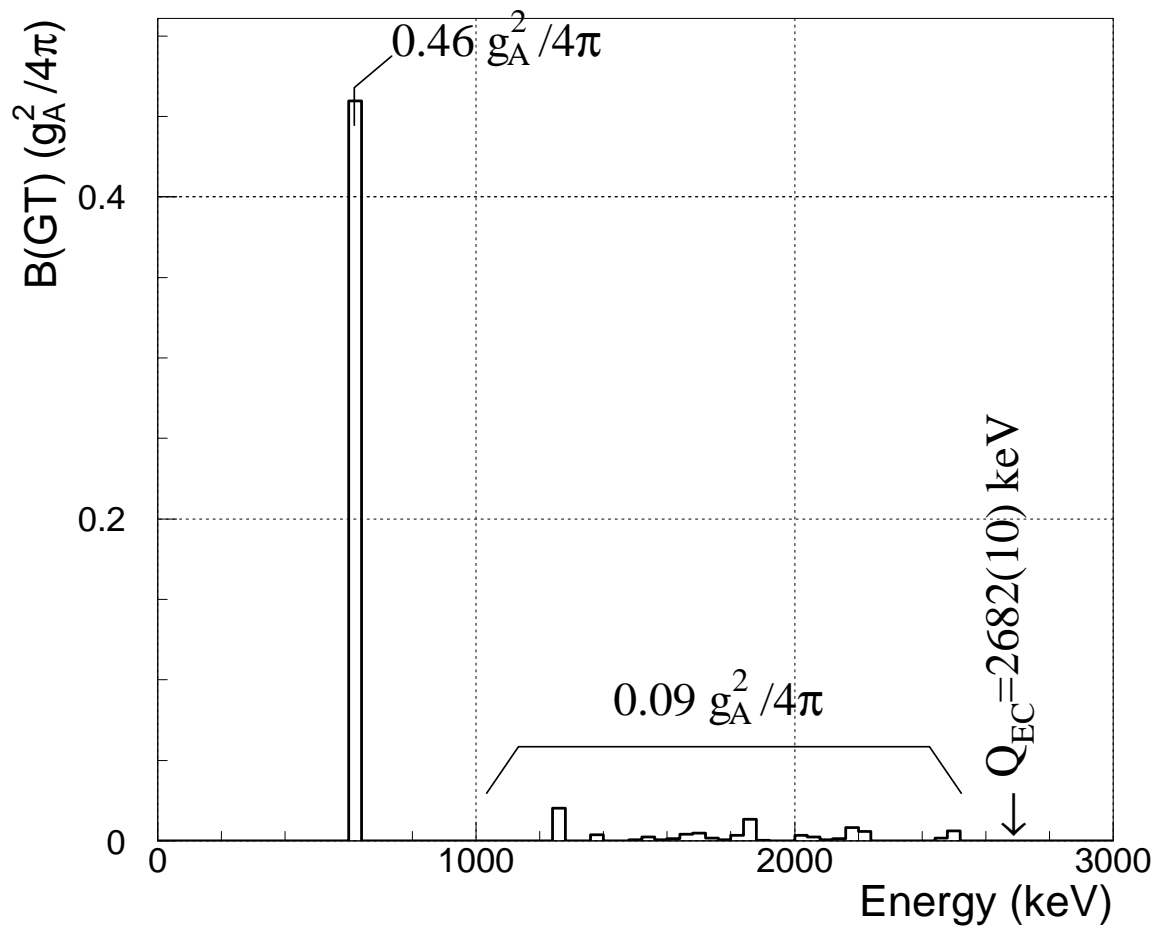

Figure 2. Gamow-Teller strength distribution in the EC/β$^+$ decay of $^{148}$Dy.

| Decay | $T_{1/2}$ | $BR_\nu$ | $EC/\nu$ | $I^\beta_{EC}$ | B(GT) | $E_{GR}$ | $\Gamma_{GR}$ | $Q_{EC}$ | $E_\nu$ | $\Delta E_\nu$ |
|---|---|---|---|---|---|---|---|---|---|---|
| $^{148}Dy \rightarrow {}^{148}Tb^*$ | 3.1 m | 1 | 0.96 | 0.96 | 0.46 | 620 | | 2682 | 2062 | |
| $^{150}Dy \rightarrow {}^{150}Tb^*$ | 7.2 m | 0.64 | 1 | 1 | 0.32 | 397 | | 1794 | 1397 | |
| $^{152}Tm2^- \rightarrow {}^{152}E_T^*$ | 8.0 s | 1 | 0.45 | 0.50 | 0.48 | 4300 | 520 | 8700 | 4400 | 520 |
| $^{150}Ho2^- \rightarrow {}^{150}Dy^*$ | 72 s | 1 | 0.77 | 0.56 | 0.25 | 4400 | 400 | 7400 | 3000 | 400 |

**Table1. Four fast decays in the rare-earth region above $^{146}$Gd leading to the giant Gamow-Teller resonance. Energies are given in keV.**

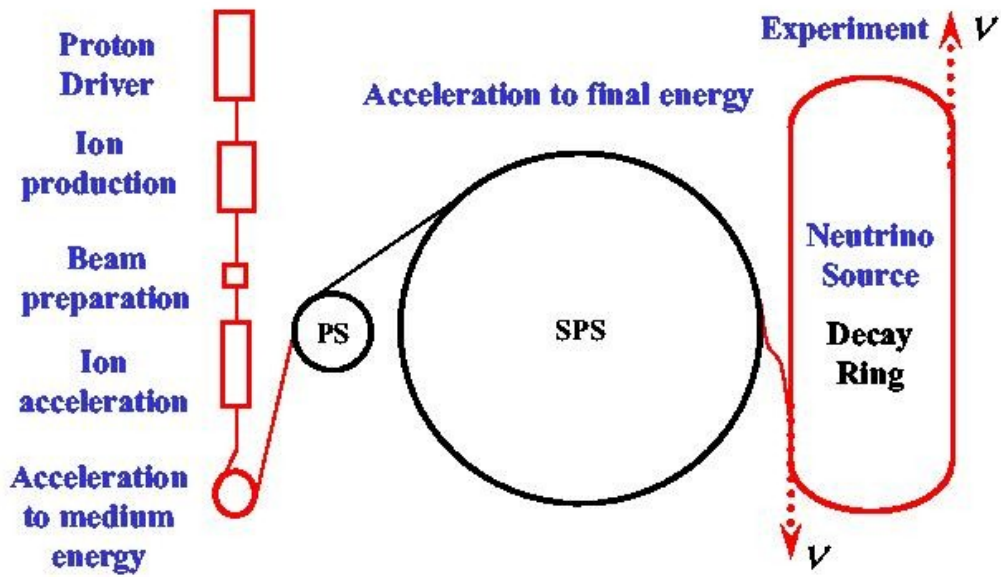

**Figure 3: A proposal for the CERN part of a "CERN to Frejus" (130 km) EC neutrino beam facility**

A neutrino (of energy $E_0$) that emerges from radioactive decay in an accelerator will be boosted in energy. At the experiment, the measured energy distribution as a function of angle (θ) and Lorentz gamma (γ) of the ion at the moment of decay can be expressed as E = $E_0$ / ($\gamma$(1- $\beta$ cos$\theta$)). The angle $\theta$ in the formula expresses the deviation between the actual neutrino detection and the ideal detector position in the prolongation of one of the long straight sections of the Decay Ring of Figure 3. The neutrinos are concentrated inside a narrow cone around the forward direction. If the ions are kept in the decay ring longer than the half-life, the energy distribution of the Neutrino Flux arriving to the detector in absence of neutrino oscillations is given by the Master Formula

$$\frac{d^2 N_\nu}{dSdE} \approx \frac{1}{\Gamma}\frac{d^2 \Gamma_\nu}{dSdE} N_{ions} \approx \frac{\Gamma_\nu}{\Gamma} \frac{N_{ions}}{\pi L^2} \gamma^2 \delta(E - 2\gamma E_0) \qquad (2)$$

with a dilation factor $\gamma \gg 1$. For an optimum choice with E ~ L around the first oscillation maximum, Eq. (2) says that lower neutrino energies $E_0$ in the proper frame give higher neutrino fluxes. The number of events will increase with higher neutrino energies as the cross section increases with energy. To conclude, in the forward direction the neutrino energy is fixed by the boost E = 2 $\gamma$ $E_0$, with the entire neutrino flux concentrated at this energy. As a result, such a facility will measure the neutrino oscillation parameters by changing the $\gamma$'s of the decay ring (energy dependent measurement) and there is no need of energy reconstruction in the detector.

We have made a simulation study in order to reach conclusions about the measurability of the unknown oscillation parameters. Some preliminary results for the Physics Reach were presented[10] before. The ion type chosen is $^{150}$Dy, with neutrino energy at rest given by 1.4 MeV due to a unique nuclear transition from 100% electron capture in going to neutrinos. Some 64% of the decay will happen as electron-capture, the rest goes through alpha decay. We have assumed that a flux of $10^{18}$/y neutrinos at the end of the long straight section of the storage ring can be obtained (e.g. at the future European nuclear physics facility, EURISOL). We have taken two energies, defined by $\gamma_{max}$ = 195 as the maximum energy possible at CERN with the present accelerator complex, and a minimum, $\gamma_{min}$ = 90, in order to avoid background in the detector below a certain energy. For the distance between source and detector we have chosen L = 130 km which equals the distance from CERN to the underground laboratory LSM in Frejus. The two values of γ are represented as vertical lines in Fig. 1. The detector has an active mass of 440 kton and the statistics is accumulated during 10 years, shared between the two runs at different γ's. The Physics Reach is represented by means of the plot in the parameters ($\theta_{13}$, δ) as given in Fig. 4, with the expected results shown as confidence level lines for the assumed values (8°, 0°), (5°, 90°), (2°, 0°) and (1°, -90°).

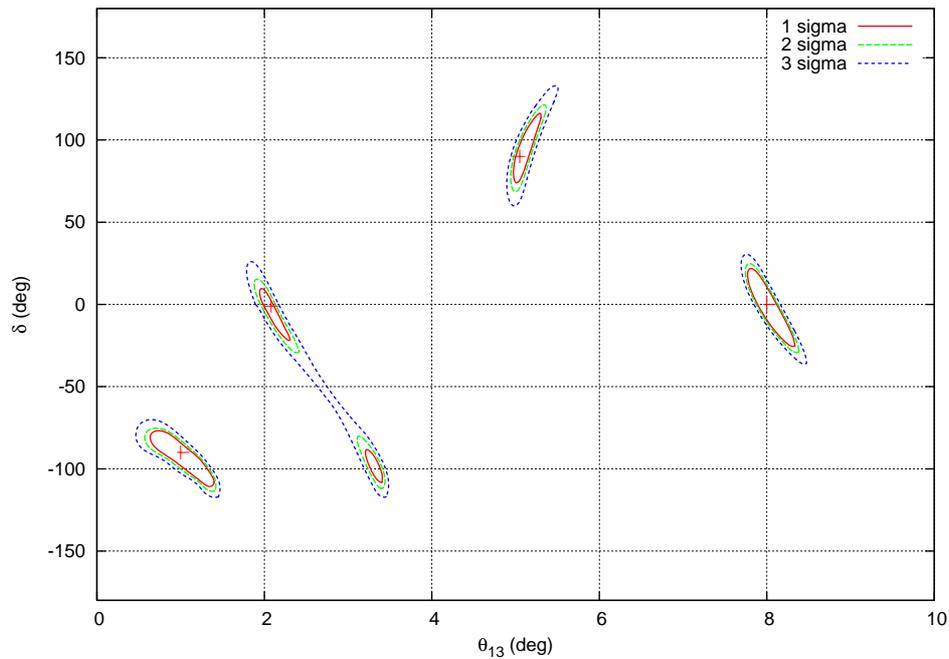

**Figure 4. Physics Reach for the presently unknown $(\theta_{13}, \delta)$ parameters, using two definite energies in the electron-capture facility discussed in this paper**.

The main conclusion is that **the principle of an energy dependent measurement is working and a window is open to the discovery of CP violation in**

**neutrino oscillations,** in spite of running at two energies only. The opportunity is better for higher values of the mixing angle $\theta_{13}$, (the angle linked to the mixing matrix element $|U_{e3}|$ and for small mixing one would need to enter into the interference region of the neutrino oscillation by going to higher distance between source and detectors.

The electron-capture facility, proposed in this work, will require a different approach to acceleration and storage of the ion beam compared to the standard beta-beam[11], as the ions cannot be fully stripped. Partly charged ions have a short vacuum life-time[12] due to a large cross-section for stripping through collisions with rest gas molecules in the accelerators. The isotopes discussed here have a half-life comparable to, or smaller than, the typical vacuum half-life of partly charged ions in an accelerator with very good vacuum. The fact that the total half-life is not dominated by vacuum losses will permit an important fraction of the stored ions sufficient time to decay through electron-capture before being lost out of the storage ring through stripping. A detailed study of production cross-sections, target and ion source designs, ion cooling and accumulation schemes, possible vacuum improvements and stacking schemes is required in order to reach a definite answer on the achievable flux. The discovery of isotopes with half-lives of a few minutes or less, which decay mainly through electron-capture to Gamow-Teller resonances in super allowed transitions, certainly opens the possibility for a monochromatic neutrino beam facility which is well worth exploring. The Physics Reach that we have shown here is impressive and demands such a study.

This research has been funded by the Grant FPA/2002-00612 and we recognize the support of the EU-I3-CARE-BENE network. We acknowledge discussions with H-C. Hseuh, M. Hjort-Jensen, E. Nacher, B. Rubio and D. Wark.

1. Y. Fukuda et al., Evidence for Oscillation of Atmospheric Neutrinos", *Phys. Rev. Lett*. **81** (1998) 1562.

2. Q. R. Ahmad et al., "Direct Evidence for Neutrino Flavor Transformation from Neutral-Current Interactions in the Sudbury Neutrino Observatory", *Phys. Rev. Lett*. **89** (2002) 011301 and 011302.

3. A. Algora et al., "β decay of $^{148}$Dy: Study of Gamow-Teller giant state by means of total absortion spectroscopy", *Phys. Rev. C* **70** (2004) 064301.

4. J. Bernabeu and M. Lindroos, "Monochromatic Neutrino Beam from Electron-Capture", in EU CARE-BENE meeting, 16-18 March 2005, CERN, Geneva, Switzerland (http://bene.na.infn.it/).

5. M. Apollonio et al., "Limits on neutrino oscillations from the CHOOZ experiment", *Phys. Lett. B* **466** (1999) 415.

6. M. C. Gonzalez-Garcia, "Global Analysis of Neutrino Data", in Nobel Symposium 2004: Neutrino Physics, Haga Slott, Enkoping, Sweden, 19-24 Aug 2004, hep-ph/0410030.

7. M. Lindroos, "Neutrino beta beam facility", in Workshop on Terrestrial and Cosmic Neutrinos, Leptogenesis and Cosmology, 4-23 July 2004, Benasque Center for Science, Spain (http://benasque.ecm.ub.es/benasque/2004neutrinos/).

8. E. Nacher, "Beta decay studies in the N≈Z and the rare-earth regions using Total Absorption Spectroscopy techniques", Ph. D. Thesis, Univ. Valencia (2004).

9. P. Zucchelli, "A novel concept for a $\bar{\nu}_e$ neutrino factory", *Phys. Lett. B* **532** (2002) 166.

10. J. Burguet-Castell, "Electron Capture", in JOINT UK Nuclear and Particle Physics meeting on the beta-beam and FIRST meeting of the EURISOL beta-beam task, 17-18 January 2005, Rutherford Appleton Laboratory, UK (http://beta-beam.web.cern.ch/beta-beam/RAL05/RAL05-presentations.htm).

11. B. Autin et al., "The acceleration and storage of radioactive ions for a neutrino factory", CERN/PS 2002-078 (OP), Proc. Nufact 02, London, UK, 2002, *J. Phys. G: Nucl. Part. Phys.* **29** (2003) 1785.

12. B. Franzke, "Vacuum Requirements for Heavy Ion Synchrotrons", Proc. 1981 Particle Accelerator Conference, *IEEE Trans. Nucl. Sci.* **NS-28** (1981) 2116.